\documentclass[runningheads]{llncs}
\usepackage{graphicx}
\usepackage[pdfborder={0 0 0.1}]{hyperref}
\usepackage{todonotes}
\usepackage{amsmath}
\usepackage{amssymb}
\usepackage{algorithm2e}
\usepackage[nocomma]{optidef}
\usepackage{cleveref}
\usepackage{enumitem}
\usepackage{graphicx}
\usepackage{booktabs}
\usepackage{multirow}
\usepackage{siunitx}
\usepackage[labelfont=bf]{caption} 
\usepackage{listings}
\usepackage{nicefrac}
\usepackage{nicematrix}

\usepackage{xurl}

\SetAlCapSkip{0pt}
\SetKwInput{KwInput}{In}
\SetKwInput{KwOutput}{Out}
\SetKwInput{KwInOut}{InOut}

\newcommand{\set}[1]{\{#1\}}

\newcommand{\dff}[1]{f_{\textsl{#1}}}

\newcommand{\lb}[1]{L_{\textsl{#1}}}
\newcommand{\bp}{\texttt{bin\_packing}}
\newcommand{\ncs}{\#\mathit{CS}}

\begin{document}
\title{Constraint Propagation on GPU: \\
A Case Study for the Bin Packing Constraint}

\titlerunning{Constraint Propagation on GPU: Bin Packing}
%
\author{
Fabio Tardivo\inst{1}\orcidID{00000-0003-3328-2174} \and
Laurent Michel\inst{2}\orcidID{0000-0001-7230-7130} \and
Enrico Pontelli \inst{1}\orcidID{0000-0002-7753-1737} 
}
\authorrunning{F. Tardivo et al.}
%
\institute{
New Mexico State University \email{\{ftardivo,epontell\}@nmsu.edu} \and
University of Connecticut \email{laurent.michel@uconn.edu}
}

\maketitle              

\begin{abstract}
The Bin Packing Problem is one of the most important problems in discrete optimization, as it captures the requirements of many real-world problems. 
Because of its importance, it has been approached with the main theoretical and practical tools. Resolution approaches based on Linear Programming are the most effective, while Constraint Programming proves valuable when the Bin Packing Problem is a component of a larger problem.

This work focuses on the Bin Packing constraint and explores how GPUs can be used to enhance its propagation algorithm. Two approaches are motivated and discussed, one based on  knapsack reasoning and one using alternative lower bounds. 
The implementations are evaluated in comparison with state-of-the-art approaches on different benchmarks from the literature. The results indicate that the GPU-accelerated lower bounds offers a  desirable alternative to tackle large instances.

\end{abstract}

\section{Introduction}
The \emph{Bin Packing Problem (BPP)} consists of packing a set of items into the minimal number of bins, each with a fixed capacity. It has a fundamental role in logistics and resource management applications, making it one of the most important optimization problems.

The BPP is NP-Hard in the strong sense \cite{Garey79} and it is challenging to solve even for a fixed number of bins \cite{Jansen13} or a constant number of different item sizes \cite{Goemans20}. 
Over the last eighty years, numerous approaches to solve the BPP have been developed; we refer the interested reader to \cite{Delorme16,Scheithauer18} for a comprehensive review. 
Techniques based on Integer Linear Programming (ILP) are highly effective and represent the state-of-the-art for solving the BPP. However, when the BPP is a component of a larger problem, applying such techniques becomes challenging, and \emph{Constraint Programming (CP)} emerges as a valuable alternative. In these cases, the BPP appears in its decision version, where the items must be packed into a fixed number of bins.

The decision version of the BPP is modeled in CP using the \bp{} constraint \cite{Shaw04}. Its filtering algorithm employs knapsack reasoning, to exclude or commit items to bins, and a feasibility check to prune the search if the remaining unpacked items cannot fit in the residual space of the bins. 
This check is performed using a lower bound on the number of bins necessary to pack the items. Typically, a combinatorial lower bound, named $L_2$, is used, but \cite{Cambazard10} has shown that employing a tighter lower bound from the linear relaxation of a strong ILP formulation, known as Arc-Flow, greatly enhances the pruning.

This work explores the use of \emph{Graphical Processing Units (GPUs)} for propagating the \bp{} constraint. The contributions of this paper include: 1) an enhanced feasibility check achieved by replacing $L_2$ with a collection of lower bounds; 2) the use of GPUs to parallelize the calculation of such lower bounds; 3) an empirical evaluation of sequential and GPU-accelerated lower bounds, compared to $L_2$ and to the lower bound from the  Arc-Flow model.

The rest of the paper is organized as follows.
\Cref{sec:background} contains some general background about Constraint Satisfaction Problems and General-purpose computing on Graphics Processing Units (GPGPU).
\Cref{sec:binpacking} summarizes related works on the \bp{} constraint.  
\Cref{sec:design} details the design and implementation of the feasibility check enhanced with the GPU-accelerated lower bounds. 
\Cref{sec:experiments} presents the results of our approach and the other techniques in the literature.
Finally, \Cref{sec:conslusions} concludes the paper.

\section{Background}
\label{sec:background}
\subsection{Constraint Satisfaction/Optimization Problems}
A Constraint Satisfaction Problem (CSP) is defined as $P = \langle V, D,C\rangle$, where $V = \set{V_1,\dots,V_n}$ is a set of \emph{variables}, $D = \set{D_1,\dots,D_n}$ is a set of \emph{domains}, and $C$ is a set of \emph{constraints}.
A constraint $c \in C$, involves a set of $m$ variables depending on its semantic. Such set is $vars(c) = \set{V_{i_1},\dots,V_{i_m}} \subseteq V$, and defines a relation $c \subseteq D_{i_1}\times\cdots\times D_{i_m}$.
A \emph{solution} is an assignment $\sigma: V \rightarrow \bigcup_{i=1}^n D_i$ such that $\sigma(V_i) \in D_i$ holds for every variable, and $\langle\sigma(V_{i_1}),\dots,\sigma(V_{i_m})\rangle \in c$ holds for every constraint.
A Constraint Optimization Problem (COP) is a quadruple $\langle {V}, {D}, {C} ,f\rangle$ where $\langle {V}, {D}, {C}\rangle$ is a CSP and  $f:{D}_{1}\times\cdots\times{D}_{n}\rightarrow \mathbb{R}$ is an \emph{objective function} to be minimized. The goal is to find a solution $\sigma$ that minimizes $f(\sigma(V_1),\cdots,\sigma(V_n))$.

A \emph{constraint solver} searches for solutions of a CSP/COP by alternating non-deterministic choices and constraints propagation. The first is employed to choose the next variable and which value, from its current domain, to assign to it. The second is a method to \emph{filter} the domain of the variables, removing values that are not part of any solution.
Non-deterministic choices are typically implemented through backtracking and heuristic decisions that follow an ordering among variables and values.
%
Constraint propagation is commonly implemented through a queue that tracks constraints that need to be re-evaluated. When a value is removed from a variable's domain, the constraints involving such variable are enqueued. The re-evaluation consists of extracting the constraint from the queue and applying the associated filtering algorithm or \emph{propagator}. This iterative cycle continues until the queue is empty \cite{Mackworth77}.

Different filtering algorithms offer different trade-offs between filtering power and computational complexity. Highly effective algorithms have been developed for \emph{global constraints}. These constraints model a substantial portion of a CSP/COP and naturally arise in various problems.

\subsection{General-Purpose Computing on Graphics Processing Units}
The computational power of modern GPUs facilitates the resolution of classes of problems that are too large to be effectively handled by CPUs. This advantage arises from GPUs massive parallelism, featuring thousands of computing units capable of efficiently processing vast amounts of data. However, to harness such computing power, it is crucial to employ approaches and algorithms that align with the underlying architecture of the GPU. Recent studies indicate that GPUs can be used for computational logic, including applications like SAT \cite{Dalpalu15,Collevati22}, ASP \cite{Dovier18,Dovier19}, and CP \cite{Tardivo23_1,Tardivo23_2}.

Most GPU-accelerated applications are developed using \emph{CUDA}\cite{Cuda}, a C/C++  API that exposes parallel computing primitives on NVIDIA GPUs. The part of an application executed by the CPU contains instructions for moving data to/from the GPU and offloading computation to the GPU. 

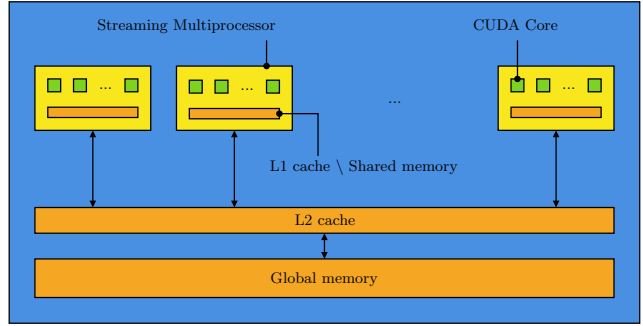
\begin{figure}[tb]
    \centering
    \resizebox{0.7\linewidth}{!}{\tikzset{every picture/.style={line width=0.75pt}} 

\begin{tikzpicture}[x=0.75pt,y=0.75pt,yscale=-1,xscale=1]

\draw  [fill={rgb, 255:red, 74; green, 144; blue, 226 }  ,fill opacity=1 ] (50,20) -- (540,20) -- (540,270) -- (50,270) -- cycle ;
\draw    (475,123) -- (475,177) ;
\draw [shift={(475,180)}, rotate = 270] [fill={rgb, 255:red, 0; green, 0; blue, 0 }  ][line width=0.08]  [draw opacity=0] (5.36,-2.57) -- (0,0) -- (5.36,2.57) -- cycle    ;
\draw [shift={(475,120)}, rotate = 90] [fill={rgb, 255:red, 0; green, 0; blue, 0 }  ][line width=0.08]  [draw opacity=0] (5.36,-2.57) -- (0,0) -- (5.36,2.57) -- cycle    ;
\draw    (225,123) -- (225,177) ;
\draw [shift={(225,180)}, rotate = 270] [fill={rgb, 255:red, 0; green, 0; blue, 0 }  ][line width=0.08]  [draw opacity=0] (5.36,-2.57) -- (0,0) -- (5.36,2.57) -- cycle    ;
\draw [shift={(225,120)}, rotate = 90] [fill={rgb, 255:red, 0; green, 0; blue, 0 }  ][line width=0.08]  [draw opacity=0] (5.36,-2.57) -- (0,0) -- (5.36,2.57) -- cycle    ;
\draw  [fill={rgb, 255:red, 248; green, 231; blue, 28 }  ,fill opacity=1 ] (70,70) -- (160,70) -- (160,120) -- (70,120) -- cycle ;
\draw  [fill={rgb, 255:red, 126; green, 211; blue, 33 }  ,fill opacity=1 ] (80,80) -- (90,80) -- (90,90) -- (80,90) -- cycle ;
\draw  [fill={rgb, 255:red, 126; green, 211; blue, 33 }  ,fill opacity=1 ] (100,80) -- (110,80) -- (110,90) -- (100,90) -- cycle ;
\draw  [fill={rgb, 255:red, 126; green, 211; blue, 33 }  ,fill opacity=1 ] (140,80) -- (150,80) -- (150,90) -- (140,90) -- cycle ;
\draw  [fill={rgb, 255:red, 245; green, 166; blue, 35 }  ,fill opacity=1 ] (80,102) -- (150,102) -- (150,110) -- (80,110) -- cycle ;
\draw  [fill={rgb, 255:red, 248; green, 231; blue, 28 }  ,fill opacity=1 ] (180,70) -- (270,70) -- (270,120) -- (180,120) -- cycle ;
\draw  [fill={rgb, 255:red, 126; green, 211; blue, 33 }  ,fill opacity=1 ] (190,81) -- (200,81) -- (200,91) -- (190,91) -- cycle ;
\draw  [fill={rgb, 255:red, 126; green, 211; blue, 33 }  ,fill opacity=1 ] (210,81) -- (220,81) -- (220,91) -- (210,91) -- cycle ;
\draw  [fill={rgb, 255:red, 126; green, 211; blue, 33 }  ,fill opacity=1 ] (250,81) -- (260,81) -- (260,91) -- (250,91) -- cycle ;
\draw  [fill={rgb, 255:red, 245; green, 166; blue, 35 }  ,fill opacity=1 ] (190,103) -- (260,103) -- (260,111) -- (190,111) -- cycle ;
\draw  [fill={rgb, 255:red, 245; green, 166; blue, 35 }  ,fill opacity=1 ] (70,180) -- (520,180) -- (520,200) -- (70,200) -- cycle ;
\draw  [fill={rgb, 255:red, 248; green, 231; blue, 28 }  ,fill opacity=1 ] (430,70) -- (520,70) -- (520,120) -- (430,120) -- cycle ;
\draw  [fill={rgb, 255:red, 126; green, 211; blue, 33 }  ,fill opacity=1 ] (440,80) -- (450,80) -- (450,90) -- (440,90) -- cycle ;
\draw  [fill={rgb, 255:red, 126; green, 211; blue, 33 }  ,fill opacity=1 ] (460,80) -- (470,80) -- (470,90) -- (460,90) -- cycle ;
\draw  [fill={rgb, 255:red, 126; green, 211; blue, 33 }  ,fill opacity=1 ] (500,80) -- (510,80) -- (510,90) -- (500,90) -- cycle ;
\draw  [fill={rgb, 255:red, 245; green, 166; blue, 35 }  ,fill opacity=1 ] (440,102) -- (510,102) -- (510,110) -- (440,110) -- cycle ;
\draw  [fill={rgb, 255:red, 245; green, 166; blue, 35 }  ,fill opacity=1 ] (70,220) -- (520,220) -- (520,250) -- (70,250) -- cycle ;
\draw    (115,123) -- (115,177) ;
\draw [shift={(115,180)}, rotate = 270] [fill={rgb, 255:red, 0; green, 0; blue, 0 }  ][line width=0.08]  [draw opacity=0] (5.36,-2.57) -- (0,0) -- (5.36,2.57) -- cycle    ;
\draw [shift={(115,120)}, rotate = 90] [fill={rgb, 255:red, 0; green, 0; blue, 0 }  ][line width=0.08]  [draw opacity=0] (5.36,-2.57) -- (0,0) -- (5.36,2.57) -- cycle    ;
\draw    (250,50) -- (250,70) ;
\draw [shift={(250,70)}, rotate = 90] [color={rgb, 255:red, 0; green, 0; blue, 0 }  ][fill={rgb, 255:red, 0; green, 0; blue, 0 }  ][line width=0.75]      (0, 0) circle [x radius= 2.01, y radius= 2.01]   ;
\draw    (445,50) -- (445,80) ;
\draw [shift={(445,80)}, rotate = 90] [color={rgb, 255:red, 0; green, 0; blue, 0 }  ][fill={rgb, 255:red, 0; green, 0; blue, 0 }  ][line width=0.75]      (0, 0) circle [x radius= 2.01, y radius= 2.01]   ;
\draw    (260,107) -- (290,107) -- (290,140) ;
\draw [shift={(260,107)}, rotate = 0] [color={rgb, 255:red, 0; green, 0; blue, 0 }  ][fill={rgb, 255:red, 0; green, 0; blue, 0 }  ][line width=0.75]      (0, 0) circle [x radius= 2.01, y radius= 2.01]   ;
\draw    (295,203) -- (295,217) ;
\draw [shift={(295,220)}, rotate = 270] [fill={rgb, 255:red, 0; green, 0; blue, 0 }  ][line width=0.08]  [draw opacity=0] (5.36,-2.57) -- (0,0) -- (5.36,2.57) -- cycle    ;
\draw [shift={(295,200)}, rotate = 90] [fill={rgb, 255:red, 0; green, 0; blue, 0 }  ][line width=0.08]  [draw opacity=0] (5.36,-2.57) -- (0,0) -- (5.36,2.57) -- cycle    ;

\draw (125,89) node [anchor=south] [inner sep=0.75pt]   [align=left] {...};
\draw (235,90) node [anchor=south] [inner sep=0.75pt]   [align=left] {...};
\draw (485,89) node [anchor=south] [inner sep=0.75pt]   [align=left] {...};
\draw (295.44,189.5) node   [align=left] {L2 cache};
\draw (294.85,235.5) node   [align=left] {Global memory};
\draw (117,32) node [anchor=north west][inner sep=0.75pt]   [align=left] {Streaming Multiprocessor};
\draw (251,141) node [anchor=north west][inner sep=0.75pt]   [align=left] {L1 cache \textbackslash \ Shared memory};
\draw (409,32) node [anchor=north west][inner sep=0.75pt]   [align=left] {CUDA Core};
\draw (349.75,100.5) node [anchor=south] [inner sep=0.75pt]   [align=left] {...};

\end{tikzpicture}}
    \caption{High level architecture of a NVIDIA GPU.}
    \label{fig:gpu_arch}
\end{figure}

The sample architecture of an NVIDIA GPU is illustrated in \Cref{fig:gpu_arch}. A modern high-end GPU is equipped with 128 \emph{Streaming Multiprocessors (SM),} each accommodating 128 computational units named \emph{CUDA Cores}. In the lower and middle tiers of the memory hierarchy, there is the \emph{global memory} with a capacity of 24 GB, and an \emph{L2 cache} of $72$ MB. At the top, there are $128$ KB of fast memory serving as \emph{L1 cache} and/or scratchpad memory (referred to as \emph{shared memory}).

The CUDA execution model is \emph{Single-Instruction Multiple-Thread} (SIMT), where a C/C++ function known as \emph{kernel} is executed by many threads. Each thread utilizes its own unique index to identify the data to use or to modify its control flow. When different threads follow distinct control flows, it leads to \emph{thread divergence}. In such scenarios, threads are serialized, potentially causing significant performance deterioration. Threads are organized into \emph{blocks}, which are dispatched to the Streaming Multiprocessors. Each Streaming Multiprocessor executes the threads using its CUDA Cores, allowing efficient intra-block operations through shared memory. Communication between blocks is possible only through the use of global memory.

To successfully GPU accelerate an application, it is crucial to achieve good load balance, optimize memory access, and mitigate thread divergence \cite{Hwu22}. This may require reformulating the problem to expose parallelism or exploiting shared memory to reduce the overhead of costly global memory accesses.

\section{Bin Packing}
\label{sec:binpacking}
Let $I = (c, W)$ be an instance of the Bin Packing Problem (BPP) with $n$ items of weights $W = [w_1, \dots, w_n]$, and bins of capacity $c$. The \emph{textbook} Integer Linear Programming (ILP) model is:
{
\small
\begin{mini*}[2]
{}{\sum_{j = 1}^{n}{y_j}}{\protect\label{model:textbook}}{}
\addConstraint{\sum_{i = 1}^{n} w_i x_{ij}} {\leq c y_j \quad} {\ \ \, j = 1,\dots,n}
\addConstraint{\sum_{j = 1}^{n} x_{ij}} {=1} {\ \ \, i = 1,\dots,n}
\addConstraint{y_j} {\in \set{0,1}} {\ \ \, j = 1,\dots,n}
\addConstraint{x_{ij}} {\in \set{0,1}} {i,j = 1,\dots,n}
\end{mini*}
}
where the Boolean variable $y_j$ indicates whether the $j^{th}$ bin is used and the variable $x_{ij}$ indicates whether
the $i^{th}$ item is packed in the $j^{th}$ bin.

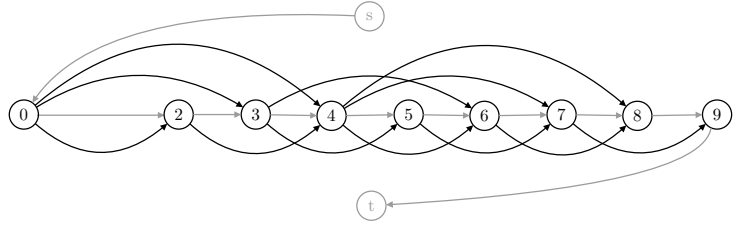
\begin{figure}[tb]
    \centering
     \resizebox{0.8\linewidth}{!}{\tikzset{every picture/.style={line width=0.75pt}} 

\begin{tikzpicture}[x=0.75pt,y=0.75pt,yscale=-1,xscale=1]

\draw [color={rgb, 255:red, 155; green, 155; blue, 155 }  ,draw opacity=1 ]   (76.93,144) -- (173.13,144) ;
\draw [shift={(176.13,144)}, rotate = 180] [fill={rgb, 255:red, 155; green, 155; blue, 155 }  ,fill opacity=1 ][line width=0.08]  [draw opacity=0] (5.36,-2.57) -- (0,0) -- (5.36,2.57) -- cycle    ;

\draw    (66.5, 143.5) circle [x radius= 11.34, y radius= 11.34]   ;
\draw (66.5,143.5) node  [font=\normalsize] [align=left] {0};
\draw    (187, 143.5) circle [x radius= 11.34, y radius= 11.34]   ;
\draw (187,143.5) node  [font=\normalsize] [align=left] {2};
\draw    (247, 143.5) circle [x radius= 11.34, y radius= 11.34]   ;
\draw (247,143.5) node  [font=\normalsize] [align=left] {3};
\draw    (306, 144.5) circle [x radius= 11.34, y radius= 11.34]   ;
\draw (306,144.5) node  [font=\normalsize] [align=left] {4};
\draw    (366, 143.5) circle [x radius= 11.34, y radius= 11.34]   ;
\draw (366,143.5) node  [font=\normalsize] [align=left] {5};
\draw    (425, 144.5) circle [x radius= 11.34, y radius= 11.34]   ;
\draw (425,144.5) node  [font=\normalsize] [align=left] {6};
\draw    (485, 143.5) circle [x radius= 11.34, y radius= 11.34]   ;
\draw (485,143.5) node  [font=\normalsize] [align=left] {7};
\draw    (544, 144.5) circle [x radius= 11.34, y radius= 11.34]   ;
\draw (544,144.5) node  [font=\normalsize] [align=left] {8};
\draw    (606, 143.5) circle [x radius= 11.34, y radius= 11.34]   ;
\draw (606,143.5) node  [font=\normalsize] [align=left] {9};
\draw  [color={rgb, 255:red, 155; green, 155; blue, 155 }  ,draw opacity=1 ]  (337, 214.5) circle [x radius= 11.34, y radius= 11.34]   ;
\draw (337,214.5) node  [font=\normalsize] [align=left] {\textcolor[rgb]{0.61,0.61,0.61}{t}};
\draw  [color={rgb, 255:red, 155; green, 155; blue, 155 }  ,draw opacity=1 ]  (335.5, 67) circle [x radius= 11.34, y radius= 11.34]   ;
\draw (335.5,67) node  [font=\normalsize] [align=left] {\textcolor[rgb]{0.61,0.61,0.61}{s}};
\draw    (76.3,137.8) .. controls (128.97,105.21) and (181.79,104.73) .. (234.77,136.36) ;
\draw [shift={(237.19,137.82)}, rotate = 211.1] [fill={rgb, 255:red, 0; green, 0; blue, 0 }  ][line width=0.08]  [draw opacity=0] (5.36,-2.57) -- (0,0) -- (5.36,2.57) -- cycle    ;
\draw    (75.26,136.31) .. controls (149.18,73.2) and (222.48,72.85) .. (295.13,135.29) ;
\draw [shift={(297.33,137.2)}, rotate = 220.96] [fill={rgb, 255:red, 0; green, 0; blue, 0 }  ][line width=0.08]  [draw opacity=0] (5.36,-2.57) -- (0,0) -- (5.36,2.57) -- cycle    ;
\draw    (195.72,150.74) .. controls (228.74,180.7) and (261.88,181.59) .. (295.14,153.38) ;
\draw [shift={(297.18,151.62)}, rotate = 138.54] [fill={rgb, 255:red, 0; green, 0; blue, 0 }  ][line width=0.08]  [draw opacity=0] (5.36,-2.57) -- (0,0) -- (5.36,2.57) -- cycle    ;
\draw    (255.78,150.67) .. controls (289.05,180.35) and (322.2,180.96) .. (355.22,152.48) ;
\draw [shift={(357.24,150.7)}, rotate = 138.06] [fill={rgb, 255:red, 0; green, 0; blue, 0 }  ][line width=0.08]  [draw opacity=0] (5.36,-2.57) -- (0,0) -- (5.36,2.57) -- cycle    ;
\draw    (256.82,137.83) .. controls (309.35,105.54) and (361.38,105.33) .. (412.92,137.2) ;
\draw [shift={(415.27,138.68)}, rotate = 212.52] [fill={rgb, 255:red, 0; green, 0; blue, 0 }  ][line width=0.08]  [draw opacity=0] (5.36,-2.57) -- (0,0) -- (5.36,2.57) -- cycle    ;
\draw    (314.83,151.61) .. controls (348.72,181.33) and (381.88,181.98) .. (414.31,153.54) ;
\draw [shift={(416.29,151.76)}, rotate = 137.56] [fill={rgb, 255:red, 0; green, 0; blue, 0 }  ][line width=0.08]  [draw opacity=0] (5.36,-2.57) -- (0,0) -- (5.36,2.57) -- cycle    ;
\draw    (315.79,138.78) .. controls (368.8,105.87) and (421.16,105.05) .. (472.85,136.33) ;
\draw [shift={(475.21,137.78)}, rotate = 211.97] [fill={rgb, 255:red, 0; green, 0; blue, 0 }  ][line width=0.08]  [draw opacity=0] (5.36,-2.57) -- (0,0) -- (5.36,2.57) -- cycle    ;
\draw    (314.78,137.33) .. controls (388.74,74.52) and (461.53,73.86) .. (533.15,135.32) ;
\draw [shift={(535.32,137.21)}, rotate = 221.22] [fill={rgb, 255:red, 0; green, 0; blue, 0 }  ][line width=0.08]  [draw opacity=0] (5.36,-2.57) -- (0,0) -- (5.36,2.57) -- cycle    ;
\draw    (374.83,150.61) .. controls (408.72,180.33) and (441.88,180.98) .. (474.31,152.54) ;
\draw [shift={(476.29,150.76)}, rotate = 137.56] [fill={rgb, 255:red, 0; green, 0; blue, 0 }  ][line width=0.08]  [draw opacity=0] (5.36,-2.57) -- (0,0) -- (5.36,2.57) -- cycle    ;
\draw    (433.82,151.63) .. controls (468.36,181.99) and (501.58,182.68) .. (533.47,153.72) ;
\draw [shift={(535.42,151.91)}, rotate = 136.52] [fill={rgb, 255:red, 0; green, 0; blue, 0 }  ][line width=0.08]  [draw opacity=0] (5.36,-2.57) -- (0,0) -- (5.36,2.57) -- cycle    ;
\draw    (493.89,150.54) .. controls (528.4,180.3) and (562.17,180.95) .. (595.21,152.47) ;
\draw [shift={(597.24,150.69)}, rotate = 138.06] [fill={rgb, 255:red, 0; green, 0; blue, 0 }  ][line width=0.08]  [draw opacity=0] (5.36,-2.57) -- (0,0) -- (5.36,2.57) -- cycle    ;
\draw [color={rgb, 255:red, 155; green, 155; blue, 155 }  ,draw opacity=1 ]   (258.33,143.69) -- (291.67,144.26) ;
\draw [shift={(294.67,144.31)}, rotate = 180.97] [fill={rgb, 255:red, 155; green, 155; blue, 155 }  ,fill opacity=1 ][line width=0.08]  [draw opacity=0] (5.36,-2.57) -- (0,0) -- (5.36,2.57) -- cycle    ;
\draw [color={rgb, 255:red, 155; green, 155; blue, 155 }  ,draw opacity=1 ]   (496.33,143.69) -- (529.67,144.26) ;
\draw [shift={(532.67,144.31)}, rotate = 180.97] [fill={rgb, 255:red, 155; green, 155; blue, 155 }  ,fill opacity=1 ][line width=0.08]  [draw opacity=0] (5.36,-2.57) -- (0,0) -- (5.36,2.57) -- cycle    ;
\draw [color={rgb, 255:red, 155; green, 155; blue, 155 }  ,draw opacity=1 ]   (601.37,153.85) .. controls (596.28,183.99) and (512.74,203.92) .. (350.76,213.66) ;
\draw [shift={(348.31,213.8)}, rotate = 356.61] [fill={rgb, 255:red, 155; green, 155; blue, 155 }  ,fill opacity=1 ][line width=0.08]  [draw opacity=0] (5.36,-2.57) -- (0,0) -- (5.36,2.57) -- cycle    ;
\draw    (75.37,150.56) .. controls (109.07,179.86) and (142.65,180.46) .. (176.1,152.34) ;
\draw [shift={(178.15,150.59)}, rotate = 138.8] [fill={rgb, 255:red, 0; green, 0; blue, 0 }  ][line width=0.08]  [draw opacity=0] (5.36,-2.57) -- (0,0) -- (5.36,2.57) -- cycle    ;
\draw [color={rgb, 255:red, 155; green, 155; blue, 155 }  ,draw opacity=1 ]   (324.17,66.57) .. controls (191.41,60.71) and (108.23,82.52) .. (74.62,132.02) ;
\draw [shift={(73.12,134.3)}, rotate = 302.52] [fill={rgb, 255:red, 155; green, 155; blue, 155 }  ,fill opacity=1 ][line width=0.08]  [draw opacity=0] (5.36,-2.57) -- (0,0) -- (5.36,2.57) -- cycle    ;
\draw [color={rgb, 255:red, 155; green, 155; blue, 155 }  ,draw opacity=1 ]   (198.34,143.5) -- (232.66,143.5) ;
\draw [shift={(235.66,143.5)}, rotate = 180] [fill={rgb, 255:red, 155; green, 155; blue, 155 }  ,fill opacity=1 ][line width=0.08]  [draw opacity=0] (5.36,-2.57) -- (0,0) -- (5.36,2.57) -- cycle    ;
\draw [color={rgb, 255:red, 155; green, 155; blue, 155 }  ,draw opacity=1 ]   (317.33,144.31) -- (351.67,143.74) ;
\draw [shift={(354.67,143.69)}, rotate = 179.05] [fill={rgb, 255:red, 155; green, 155; blue, 155 }  ,fill opacity=1 ][line width=0.08]  [draw opacity=0] (5.36,-2.57) -- (0,0) -- (5.36,2.57) -- cycle    ;
\draw [color={rgb, 255:red, 155; green, 155; blue, 155 }  ,draw opacity=1 ]   (377.33,143.69) -- (410.67,144.26) ;
\draw [shift={(413.67,144.31)}, rotate = 180.97] [fill={rgb, 255:red, 155; green, 155; blue, 155 }  ,fill opacity=1 ][line width=0.08]  [draw opacity=0] (5.36,-2.57) -- (0,0) -- (5.36,2.57) -- cycle    ;
\draw [color={rgb, 255:red, 155; green, 155; blue, 155 }  ,draw opacity=1 ]   (436.33,144.31) -- (470.67,143.74) ;
\draw [shift={(473.67,143.69)}, rotate = 179.05] [fill={rgb, 255:red, 155; green, 155; blue, 155 }  ,fill opacity=1 ][line width=0.08]  [draw opacity=0] (5.36,-2.57) -- (0,0) -- (5.36,2.57) -- cycle    ;
\draw [color={rgb, 255:red, 155; green, 155; blue, 155 }  ,draw opacity=1 ]   (555.33,144.32) -- (591.67,143.73) ;
\draw [shift={(594.67,143.68)}, rotate = 179.08] [fill={rgb, 255:red, 155; green, 155; blue, 155 }  ,fill opacity=1 ][line width=0.08]  [draw opacity=0] (5.36,-2.57) -- (0,0) -- (5.36,2.57) -- cycle    ;

\end{tikzpicture}}
    \caption{Graph underling the Arc-Flow model for an instance with  $c = 9$ and $W = [4, 4, 3, 3, 2, 2]$.}
    \label{fig:arc-flow}
\end{figure}

A strong ILP formulation, known as \emph{Arc-Flow} \cite{Carvalho99}, is obtained by approaching the BPP from a graph-theoretical perspective. Given a BPP instance, a graph is built in such a way that arcs represent items, and a path from the source node $s$ to the sink node $t$ represents a set of items that can be packed into a bin (see \cref{fig:arc-flow}). A solution corresponds to a minimum flow that uses one arc for each $w \in W$. The ILP formulation of this flow problem has a strong linear relaxation, but it comes at the cost of a pseudo-polynomial number of variables and constraints.

In CP, the decision version of the BPP, where the items must be packed in at most $k$ bins, is modeled as:
\begin{gather*}
    x_i = \{1,\dots,k\} \qquad i = 1, \dots, n\\
    l_j = \{0,\dots,c\} \qquad j = 1, \dots, k\\
    \bp([x_1, \dots, x_n], [w_1, \dots, w_n],[l_1, \dots, l_k])
\end{gather*}
where the variable $x_i$ represents the bins in which the $i^{th}$ item can be packed, and the variable $l_j$ represents the loads that the $j^{th}$ bin can have. 

The \bp{} constraint was introduced in \cite{Shaw04} and a simplified version of its filtering algorithm is listed in \cref{algo:filtering_bpp}. Following a brief description of each method:

\begin{algorithm}[tb]
    \SetAlgoVlined
    \DontPrintSemicolon
    \LinesNumbered
    \KwInput{$c$, $W = [w_1, \dots, w_n]$, $k$}
    \KwInOut{$X = [x_1, \dots, x_n]$, $L = [l_1,\dots, l_k]$}
    \For(\tcp*[f]{Basic filtering}){$j \gets 1$ \KwTo $k$}
    {
        $\mathit{doLoadCoherence}(j,X,W,L)$\;
        $\mathit{doBasicLoadTightening}(j,X,W,L)$\;
        \For{$i \in \set{i \mid j \in x_i \land \left|x_i\right| > 1}$}
        {
            $\mathit{doBasicItemEliminationCommitment}(i,j,X,W,L)$\;
        }
    }
    \For(\tcp*[f]{Knapsack filtering}){$j \gets 1$ \KwTo $k$} 
    {
        \If{$\lnot \mathit{isBinPackable}(j,X,W,L)$}
        {
           $\mathit{Fail}$\;
        }
        $\mathit{doKnapsackLoadTightening}(j,X,W,L)$\;  
        \For{$i \in \set{i \mid j \in x_i \land \left|x_i\right| > 1}$}
        {
            $\mathit{doKnapsackItemEliminationCommitment}(i,j,X,W,L)$\;
        }
    }  
    $lb \gets\mathit{getLowerBound}(c,W,k,X)$  \tcp*{Feasibility check}
    \If{$ lb > k$}
    {
        $\mathit{Fail}$\;
    }
    \caption{Simplified propagator for the \bp{} constraint.}
    \label{algo:filtering_bpp}
\end{algorithm}


\begin{description}
    \item[Load coherence] The minimum/maximum load of a bin is adjusted considering the total weight of the items and the load of the other bins.
    \item[Basic load tightening] The minimum/maximum load of a bin is adjusted considering the sum of the items that are/can be packed in the bin.
    \item[Basic item elimination and commitment] An item is committed to a bin if it is needed to reach a valid load. An item is excluded from a bin if packing it would lead to an excessive load.
    \item[Bin packability check] A bin is considered packable if an approximated knapsack reasoning shows that it is possible to reach an admissible load.
    \item[Knapsack load tightening] The minimum/maximum load of a bin is adjusted using an approximated knapsack reasoning.
    \item[Knapsack item elimination and commitment] An item is committed or excluded from a bin using an approximated knapsack reasoning.
    \item[Feasibility check] A partial packing is considered feasible if a lower bound on the number of bins does not exceed the available bins. The lower bound is calculated on a \emph{reduced} instance derived from the current partial packing (see \cref{fig:red}). This instance is obtained by considering all the unpacked items and introducing one \emph{virtual item} per bin to represent the items packed in that bin. Typically, the lower bound $L_2$ is used (see \cref{sec:lowerbounds}).
\end{description}
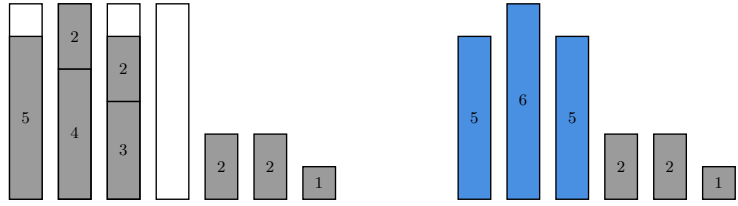
\begin{figure}[tb]
    ~\hfill
    \resizebox{4.5cm}{!}{\tikzset{every picture/.style={line width=0.75pt}} 

\begin{tikzpicture}[x=0.75pt,y=0.75pt,yscale=-1,xscale=1]

\draw   (60,96) -- (84,96) -- (84,240) -- (60,240) -- cycle ;
\draw  [fill={rgb, 255:red, 155; green, 155; blue, 155 }  ,fill opacity=1 ] (276,216) -- (300,216) -- (300,240) -- (276,240) -- cycle ;

\draw  [fill={rgb, 255:red, 155; green, 155; blue, 155 }  ,fill opacity=1 ] (96,144) -- (120,144) -- (120,240) -- (96,240) -- cycle ;

\draw   (96,96) -- (120,96) -- (120,240) -- (96,240) -- cycle ;
\draw  [fill={rgb, 255:red, 155; green, 155; blue, 155 }  ,fill opacity=1 ] (132,120) -- (156,120) -- (156,168) -- (132,168) -- cycle ;

\draw  [fill={rgb, 255:red, 155; green, 155; blue, 155 }  ,fill opacity=1 ] (132,168) -- (156,168) -- (156,240) -- (132,240) -- cycle ;

\draw   (132,96) -- (156,96) -- (156,240) -- (132,240) -- cycle ;
\draw   (168,96) -- (192,96) -- (192,240) -- (168,240) -- cycle ;
\draw  [fill={rgb, 255:red, 155; green, 155; blue, 155 }  ,fill opacity=1 ] (96,96) -- (120,96) -- (120,144) -- (96,144) -- cycle ;

\draw  [fill={rgb, 255:red, 155; green, 155; blue, 155 }  ,fill opacity=1 ] (60,120) -- (84,120) -- (84,240) -- (60,240) -- cycle ;

\draw  [fill={rgb, 255:red, 155; green, 155; blue, 155 }  ,fill opacity=1 ] (204,192) -- (228,192) -- (228,240) -- (204,240) -- cycle ;

\draw  [fill={rgb, 255:red, 155; green, 155; blue, 155 }  ,fill opacity=1 ] (240,192) -- (264,192) -- (264,240) -- (240,240) -- cycle ;

\draw (287.85,228.19) node    {$1$};
\draw (143.85,144.19) node    {$2$};
\draw (143.85,203.19) node    {$3$};
\draw (108,120) node    {$2$};
\draw (215.85,216.19) node    {$2$};
\draw (251.85,216.19) node    {$2$};
\draw (107.85,191.19) node    {$4$};
\draw (72,180) node    {$5$};

\end{tikzpicture}}
    ~\hfill~
    \resizebox{3.85cm}{!}{\tikzset{every picture/.style={line width=0.75pt}} 

\begin{tikzpicture}[x=0.75pt,y=0.75pt,yscale=-1,xscale=1]

\draw  [fill={rgb, 255:red, 155; green, 155; blue, 155 }  ,fill opacity=1 ] (348,180) -- (372,180) -- (372,204) -- (348,204) -- cycle ;

\draw  [fill={rgb, 255:red, 74; green, 144; blue, 226 }  ,fill opacity=1 ] (204,60) -- (228,60) -- (228,204) -- (204,204) -- cycle ;
\draw  [fill={rgb, 255:red, 155; green, 155; blue, 155 }  ,fill opacity=1 ] (312,156) -- (336,156) -- (336,204) -- (312,204) -- cycle ;

\draw  [fill={rgb, 255:red, 74; green, 144; blue, 226 }  ,fill opacity=1 ] (168,84) -- (192,84) -- (192,204) -- (168,204) -- cycle ;

\draw  [fill={rgb, 255:red, 155; green, 155; blue, 155 }  ,fill opacity=1 ] (276,156) -- (300,156) -- (300,204) -- (276,204) -- cycle ;

\draw  [fill={rgb, 255:red, 74; green, 144; blue, 226 }  ,fill opacity=1 ] (240,84) -- (264,84) -- (264,204) -- (240,204) -- cycle ;

\draw (215.85,130.78) node    {$6$};
\draw (252,144) node    {$5$};
\draw (287.85,180.19) node    {$2$};
\draw (180,144) node    {$5$};
\draw (323.85,180.19) node    {$2$};
\draw (359.85,192.19) node    {$1$};

\end{tikzpicture}}
    \hfill~
    \caption{Illustration of a partial packing (left) and its reduction (right). Virtual items are highlighted in blue.}
    \label{fig:red}
\end{figure}
The literature contains various extensions of the \bp{} constraint. The authors of \cite{Schaus12,Pelsser13,Derval18} introduce and refine a \emph{cardinality reasoning}, well suited when there are constraints on the number of items in each bin or when the items have similar weights. In \cite{Dupuis10}, the lower bound is enhanced by considering \emph{alternative reductions}. Finally, \cite{Cambazard10} achieves notable results using the lower bound derived from the linear relaxation of the \emph{Arc-Flow} model.

\subsection{Lower bounds}
\label{sec:lowerbounds}
Given an instance $I = (c, W)$ of the BPP, a lower bound $L(I)$ estimates the minimum number of bins necessary to store the items. The simplest lower bound is referred to as $L_1$, and  is calculated as follows:
\begin{equation*}
L_1(I) = \left\lceil \frac{1}{c} \sum_{w \in W} w \right\rceil
\end{equation*}
where the total weight of the items is divided by the bin capacity, and the ceiling function is applied. This approach is equivalent to naively packing the items, cutting those that do not entirely fit.

An improvement of $L_1$, called  $L_2$,  is introduced in \cite{Martello90} and addresses the cases where big items cannot be packed together. It is defined as:
\begin{equation*}
    L_2(I) = \max_{0 \leq \lambda \leq \frac{c}{2}} L_2(I,\lambda)
\end{equation*}

\pagebreak\noindent%
where
    \begin{equation*}
     L_2(I, \lambda) = \left|W_1\right| + \left|W_2\right| + max\left(0,\left\lceil \frac{1}{c}\left(\sum_{w  \in W_3} w -\left(c \left|W_2\right| - \sum_{w  \in W_2} w \right)\right)\right\rceil\right)
     \end{equation*}
     \begin{equation*}
     W_1 = \set{w \mid w \in W \land c - \lambda < w}\\
     \end{equation*}
     \begin{equation*}
     W_2 = \set{w \mid w \in W \land \tfrac{c}{2} < w \leq c - \lambda}\\
     \end{equation*}
     \begin{equation*}
     W_3 = \set{w \mid w \in W \land \lambda \leq w \leq \tfrac{c}{2}}
    \end{equation*}
The lower bound $L_2(I, \lambda)$ is equivalent to first classifying the items as big ($W_1$), medium-big ($W_2$), and medium-small ($W_3$), while smaller items are ignored. Each of the big and medium-big items is packed in a different bin, since they are bigger than $\frac{c}{2}$. Finally, the medium-small items are packed as in $L_1$, using the available space in the bins where there is a medium-big item before considering other bins. A direct implementation of $L_2$ is pseudo-polynomial, since $L_2(I,\lambda)$ has to be calculated $O(c)$ times. More efficient algorithms are described in \cite{Martello90_2,Korf02}, achieving linear complexity when the items are sorted in decreasing weight.

A general approach to enhance $L_1$ is based on \emph{Dual Feasible Functions} (DFFs). These functions alter the weights of items. 
In the example below, $f_{MT}(w,\lambda)$ will either increase (when $w > c -\lambda$), decrease (when $w < \lambda$) or not change (when $\lambda \leq w \leq c - \lambda$) the original weight $w$. Note how increasing the weight to be $c$ requires a dedicated bin for the item, while decreasing its weight to $0$ says the item is ignored.
%
The function is illustrated in \cref{fig:f_mt} and depends on an \emph{integer} parameter $\lambda$:
\begin{figure}[tb]
        \centering
         \resizebox{0.5\linewidth}{!}{\tikzset{every picture/.style={line width=0.75pt}} 

\begin{tikzpicture}[x=0.75pt,y=0.75pt,yscale=-1,xscale=1]

\draw    (90,200) -- (277,200) ;
\draw [shift={(280,200)}, rotate = 180] [fill={rgb, 255:red, 0; green, 0; blue, 0 }  ][line width=0.08]  [draw opacity=0] (5.36,-2.57) -- (0,0) -- (5.36,2.57) -- cycle    ;
\draw    (100,210) -- (100,23) ;
\draw [shift={(100,20)}, rotate = 90] [fill={rgb, 255:red, 0; green, 0; blue, 0 }  ][line width=0.08]  [draw opacity=0] (5.36,-2.57) -- (0,0) -- (5.36,2.57) -- cycle    ;
\draw [color={rgb, 255:red, 0; green, 0; blue, 0 }  ,draw opacity=0.1 ]   (100,200) -- (250,50) ;
\draw [line width=1.5]    (140,160) -- (210,90) ;
\draw [shift={(210,90)}, rotate = 315] [color={rgb, 255:red, 0; green, 0; blue, 0 }  ][fill={rgb, 255:red, 0; green, 0; blue, 0 }  ][line width=1.5]      (0, 0) circle [x radius= 2.61, y radius= 2.61]   ;
\draw [shift={(140,160)}, rotate = 315] [color={rgb, 255:red, 0; green, 0; blue, 0 }  ][fill={rgb, 255:red, 0; green, 0; blue, 0 }  ][line width=1.5]      (0, 0) circle [x radius= 2.61, y radius= 2.61]   ;
\draw [color={rgb, 255:red, 126; green, 211; blue, 33 }  ,draw opacity=1 ][fill={rgb, 255:red, 65; green, 117; blue, 5 }  ,fill opacity=1 ][line width=1.5]    (211.61,50) -- (250,50) ;
\draw [shift={(210,50)}, rotate = 0] [color={rgb, 255:red, 126; green, 211; blue, 33 }  ,draw opacity=1 ][line width=1.5]      (0, 0) circle [x radius= 2.61, y radius= 2.61]   ;
\draw    (250,200) -- (250,205) ;
\draw    (140,200) -- (140,205) ;
\draw [color={rgb, 255:red, 0; green, 0; blue, 0 }  ,draw opacity=0.1 ]   (250,200) -- (250,50) ;
\draw [color={rgb, 255:red, 0; green, 0; blue, 0 }  ,draw opacity=0.1 ]   (210,200) -- (210,90) ;
\draw [color={rgb, 255:red, 0; green, 0; blue, 0 }  ,draw opacity=0.1 ]   (140,200) -- (140,160) ;
\draw    (210,200) -- (210,205) ;
\draw [color={rgb, 255:red, 208; green, 2; blue, 27 }  ,draw opacity=1 ][fill={rgb, 255:red, 208; green, 2; blue, 27 }  ,fill opacity=1 ][line width=1.5]    (100,200) -- (138.39,200) ;
\draw [shift={(140,200)}, rotate = 0] [color={rgb, 255:red, 208; green, 2; blue, 27 }  ,draw opacity=1 ][fill=white][line width=1.5]      (0, 0) circle [x radius= 2.61, y radius= 2.61]   ;

\draw (250.22,219.2) node  [font=\normalsize]  {$c$};
\draw (140,219) node  [font=\normalsize]  {$\lambda $};
\draw (210.22,219.2) node  [font=\normalsize]  {$c - \lambda$};
\draw (273,215) node  [font=\normalsize]  {$w$};
\draw (62,30) node  [font=\normalsize]  {$\dff{MT}(w,\lambda )$};

\end{tikzpicture}}
        \caption{Illustration of $\dff{MT}$ for $\lambda = c\frac{4}{15}$. Weights that have been increased/decreased are highlighted in green/red.}
        \label{fig:f_mt}
\end{figure}
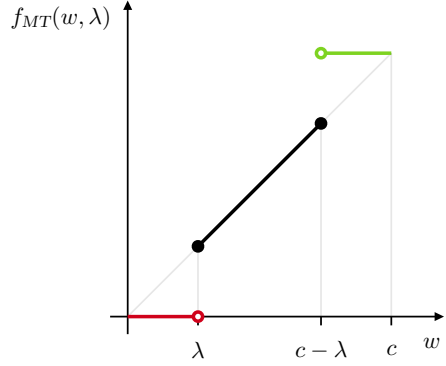
\begin{equation*}
    \underset{0 \leq \lambda \leq \frac{c}{2}}{\dff{MT}(w, \lambda)} =
    \begin{cases}
        c & \text{if $c - \lambda < w$} \\
        w & \text{if $\lambda \leq w \leq c - \lambda$} \\
        0 & \text{if $w < \lambda$} \\   
    \end{cases}
\end{equation*}
The lower bound obtained by combining $L_1$ with $\dff{MT}$ is: 
    \begin{equation*}
       \lb{MT}(I) = \max_{0 \leq \lambda \leq \frac{c}{2}} \left\lceil \frac{1}{\dff{MT}(c,\lambda)} \sum_{w \in W} \dff{MT}(w, \lambda) \right\rceil
    \end{equation*}
and it is equivalent to $L_2$ \cite{Fekete01}. Several other DFFs have been proposed, each with a different design. For brevity, we report only some of them and refer interested readers to \cite{Clautiaux10,Alves16} for a comprehensive review, and to \cite{Rietz10,Rietz12} for further insights.
\begin{align*}
    &\underset{\frac{c}{4} < \lambda \leq \frac{c}{3}}{\dff{RAD2}(w, \lambda)} =
    \begin{cases}
        0 & \text{if $w < \lambda$} \\
        \left\lfloor\frac{c}{3}\right\rfloor & \text{if $\lambda \leq w \leq c - 2\lambda$}\\
        \left\lfloor\frac{c}{2}\right\rfloor & \text{if $c - 2\lambda < w < 2\lambda$}\\
        c - \dff{RAD2}(c - w, \lambda) & \text{if $2\lambda \leq w$}\\ 
    \end{cases}\\[2mm]
    &\underset{1 \leq \lambda \leq 100}{\dff{FS1}(w, \lambda)} =
    \begin{cases}
        w\lambda & \text{if $\frac{w(\lambda + 1)}{c} \in \mathbb{N}$} \\
        \left\lfloor \frac{w(\lambda + 1)}{c} \right\rfloor c & \text{otherwise}
    \end{cases}\\[2mm]
    &\underset{1 \leq \lambda \leq \frac{c}{2}}{\dff{CCM1}(w, \lambda)} =
    \begin{cases}
        2\left\lfloor\frac{c}{\lambda}\right\rfloor - 2\left\lfloor\frac{c - w}{\lambda}\right\rfloor & \text{if $w > \frac{c}{2}$} \\
        \left\lfloor\frac{c}{\lambda}\right\rfloor  & \text{if $w = \frac{c}{2}$}\\
        2\left\lfloor\frac{w}{\lambda}\right\rfloor & \text{if $w < \frac{c}{2}$}\\
    \end{cases}\\[2mm]
    &\underset{2 \leq \lambda \leq c}{\dff{VB2}(w, \lambda)} =
    \begin{cases}
        2\max\left(0, \left\lceil \frac{c\lambda}{c} \right\rceil - 1\right) - 2 \max\left(0, \left\lceil \frac{c\lambda - w\lambda}{c} \right\rceil - 1\right) & \text{if $w > \frac{c}{2}$} \\
        \max\left(0, \left\lceil \frac{c\lambda}{c} \right\rceil - 1\right) & \text{if $w = \frac{c}{2}$}\\
        2\max\left(0, \left\lceil \frac{w\lambda}{c} \right\rceil - 1\right)                & \text{if $w < \frac{c}{2}$}\\
    \end{cases}\\[2mm]
    &\underset{1 \leq \lambda \leq c}{\dff{BJ1}(w, \lambda)} =
    \begin{cases}
        \left\lfloor\frac{w}{\lambda}\right\rfloor(\lambda - c \bmod  \lambda) & \text{if $w \bmod \lambda \leq c \bmod \lambda$} \\
        \left\lfloor\frac{w}{\lambda}\right\rfloor(\lambda - c \bmod  \lambda) + w \bmod \lambda - c \bmod \lambda & \text{otherwise}\\
    \end{cases}
\end{align*}

\section{Design and Implementation}
\label{sec:design}
A strategy for leveraging the GPU in constraint propagation involves utilizing it for \emph{strong filtering} at a reduced computational cost \cite{Tardivo23_2}. This approach can be extended to the \bp{} constraint by employing the GPU to perform a \emph{complete knapsack reasoning} instead of an approximated one.
With the exception of load coherence, all the basic and knapsack filterings in \cref{algo:filtering_bpp} can be performed using the Dynamic Programming (DP) approach presented in \cite{Trick03}. We have developed a GPU-accelerated implementation of this method, leveraging bitwise operations and processing each bin in parallel.
Initial tests did not reveal significant differences
in terms of explored nodes compared to the approximated reasoning. Scalability tests indicate that the GPU-accelerated implementation becomes faster than an optimized implementation of the approximated filtering when the number of bins is in the order of hundreds. 
Although the underlying DP tables are calculated very efficiently, this approach is hindered by the overhead resulting from the transfer of the variables to/from the GPU, the identification of the items that can be packed in each bin, and the atomic commit/elimination of the items. These results lead us to discard this approach and stick with the standard approximated knapsack reasoning.

\begin{figure}[!b]
    \centering
    \hspace*{-3.5cm}
    \resizebox{1.2\linewidth}{!}{\input{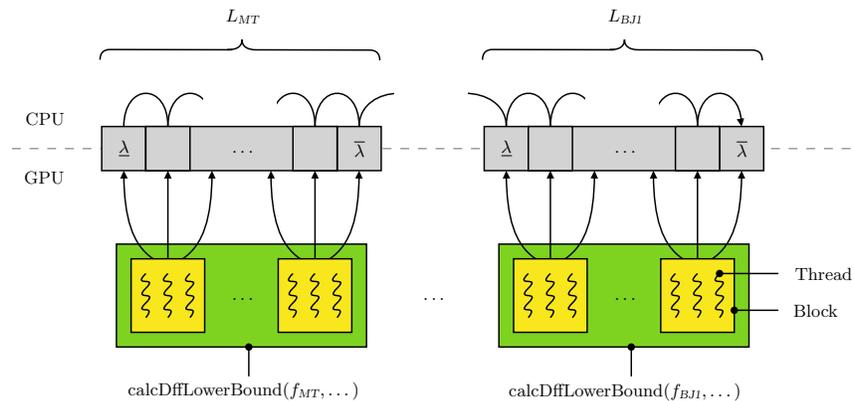}}
    \caption{Sequential (top) and parallel (bottom) execution of the DFFs-based $\mathit{getLowerBound}$ functions.}
    \label{fig:lb_par}
\end{figure}

Another strategy to leverage the GPU in constraint propagation is to employ it to enhance the \emph{pruning}. This translates into improving the feasibility check by utilizing the GPU to obtain the tightest lower bound at a reduced computational cost.
Since the best lower bound is derived from a linear relaxation, it would reduce to solving a sparse linear system, a task notoriously hard to effectively accelerate by the GPU \cite{Hwu22}.
The next tightest lower bounds are obtained using Dual Feasible Functions (DFFs), and as demonstrated later in this section, these bounds are well-suited for GPU acceleration.
The feasibility check can be easily adapted to make use of the DFFs-based lower bound listed in \cref{algo:lb_dffs}.
\begin{algorithm}[t]
    \footnotesize
    \SetAlgoVlined
    \DontPrintSemicolon
    \LinesNumbered
    \KwInput{$c$, $W = [w_1, \dots, w_n]$, $k$, $X = [x_1, \dots, x_n]$}
    \KwOutput{$lb$}
    $[w'_1, \dots, w'_r] \gets \mathit{getReduction}(W,X)$\;
    $W_R \gets [w'_1, \dots, w'_r]$\;
    $lb \gets 0$\;
    \For{$f \in \set{\dff{MT}, \dff{RAD2}, \dff{FS1}, \dff{CCM1}, \dff{VB2}, \dff{BJ1}}$} 
    {
        {
            $L_f \gets 0$\;
            $\underline{\lambda}, \overline{\lambda}  \gets \mathit{getMinMaxParameter}(f,c)$\;           
            \For{$\lambda \gets \underline{\lambda}$ \KwTo $ \overline{\lambda}$} 
            {
                $sum \gets 0$ \tcp*{Calculate $\lb{1}$-like lower bound}
                \For{$w \in W_R$}
                {
                    $sum \gets sum  + f(w, \lambda)$\;
                }
                $lb' \gets \left\lceil \frac{sum}{f(c, \lambda)} \right\rceil$\;
                $L_f \gets max(L_f, lb')$
             }
         }
          $lb \gets max(lb, L_f)$\;
          \If{$lb > k$}
          {
                \KwRet{$lb$}
          }
     }
    \KwRet{$lb$}
    \caption{Sequential DFFs-based $\mathit{getLowerBound}$ function.}
    \label{algo:lb_dffs}
\end{algorithm}
The order in which the DFFs are considered impacts how quickly an infeasible partial packing is detected, and DFFs leading to a generally good lower bound should be prioritized. \Cref{tab:dff_lb} presents a summary of the lower bounds derived from different DFFs on the Falkenauer and Scholl instances (see \cref{sec:experiments}). The column `Only Opt'/`Only Best' indicates the number of instances for which the DFF was the only one to lead to the optimal/best lower bound, while the `Sum' column represents the sum of all the lower bounds calculated from the DFF. The optimal number of bins was found for 1305 out of 1370 total instances. The results confirm $\dff{CCM1}$ as the best overall function \cite{Clautiaux10}. Interestingly, the generally weak $\dff{RAD2}$ proves effective when stronger functions are suboptimal \cite{Rietz10}.
\begin{table}[t]
    \footnotesize
    \setlength{\tabcolsep}{6pt} 
    \centering
    \begin{tabular}{@{}crrrrr@{}}
        DFF        & Only Opt & Total Opt & Only Best & Total Best & Sum    \\ 
        \midrule
        $\dff{MT}$   & 2       & 1151     & 0        & 55   & 120184 \\   
        $\dff{RAD2}$ & 10      & 189      & 0        & 36   & 105345 \\ 
        $\dff{FS1}$  & 2       & 742      & 0        & 45   & 119504 \\
        $\dff{CCM1}$ & 40      & 1219     & 1        & 60   & 120270 \\
        $\dff{VB2}$  & 1       & 973      & 0        & 40   & 119786 \\
        $\dff{BJ1}$  & 47      & 1101     & 0        & 50   & 120039 \\
    \end{tabular}
    \caption{Statistics of different DFF-based lower bounds.}
    \label{tab:dff_lb}
\end{table}
\paragraph{Parallelization} %
The GPU-accelerated feasibility check is outlined in \cref{algo:lb_dffs_par}.
The core of the parallelization is a kernel named \emph{calcDffLowerBound}, which is responsible for calculating the lower bounds derived from the different parameters of a DFF and keeping track of the tightest one (see \cref{algo:lb_dffs_kernel}).
Concurrently running different copies of the kernel, each working with a different DFF, parallelizes the first outermost loop of \cref{algo:lb_dffs}. Using the threads to calculate the lower bounds parallelizes the second outermost loop (see \cref{fig:lb_par}).
Each thread is responsible for a distinct parameter value, requiring that $nThreads \geq \overline{\lambda} - \underline{\lambda} + 1$. 
This condition is satisfied by launching each kernel with $\left\lceil\frac{\overline{\lambda} - \underline{\lambda} + 1}{\ncs}\right\rceil$ blocks of size $\ncs$, where $\ncs$ is the number of CUDA Cores per Streaming Multiprocessors.  With a bin capacity of $2000$, a current high-end GPU  achieves full parallelization of up to 8 DFFs, with a linear performance penalty for each additional DFF.

\paragraph{Implementation details} The \emph{calcDffLowerBound} kernel includes a couple of memory optimizations not represented in the pseudocode. First, each block initially copies $[w'_1, \dots, w'_r]$ into shared memory, ensuring fast accessibility for the subsequent lower bounds calculations. Second, the shared memory is also used to store $L_f$, allowing faster atomic operations that can run concurrently between blocks and reducing the number of atomic operations performed in the slower global memory.

Another optimization is employed at a higher level and consists of batching the memory operations and kernel launches to be called using a \emph{single} API call. Depending on the instance, this technique provides a speedup up to 2x.

Finally, a note on numeric overflows. When the capacity is large, the intermediate values in the lower bound derived from $\dff{VB2}$ can exceed \texttt{UINT\_MAX}.
To address this problem, we limit the maximum parameter of $\dff{VB2}$ to $\left\lfloor\frac{\texttt{UINT\_MAX}}{r\cdot max(w'_1, \dots, w'_r)}\right\rfloor$.
\begin{algorithm}[!t]
    \footnotesize
    \SetAlgoVlined
    \DontPrintSemicolon
    \LinesNumbered
    \KwInput{$c$, $W = [w_1, \dots, w_n]$, $k$, $X = [x_1, \dots, x_n]$}
    \KwOutput{$lb$}
    $[w'_1, \dots, w'_r] \gets \mathit{getReduction}(W,X)$\;
    $W_R \gets [w'_1, \dots, w'_r]$\;
    $lb \gets 0$\;
    $\mathit{cudaMemcpyCpuToGpu}(c, W_R, lb)$ \tcp*{Asynchronous APIs}
    $\underline{\lambda}, \overline{\lambda}  \gets \mathit{getMinMaxParameter}(\dff{MT},c)$\;
    $\mathit{cudaLaunchKernel}(\mathit{calcDffLowerBound},\mathit{nThreads}, [\dff{MT}, \underline{\lambda}, \overline{\lambda}, c, W_R, k, lb])$\;
    $\cdots$\;
    $\underline{\lambda}, \overline{\lambda}  \gets \mathit{getMinMaxParameter}(\dff{BJ1},c)$\;
     $\mathit{cudaLaunchKernel}(\mathit{calcDffLowerBound},\mathit{nThreads}, [\dff{BJ1}, \underline{\lambda}, \overline{\lambda}, c, W_R,k,lb])$\;
    $\mathit{cudaMemcpyGpuToCpu(lb)}$\;
    $\mathit{waitGpu}()$ \tcp*{Synchronous API}
    \KwRet{$lb$}
    \caption{Parallel DFFs-based $\mathit{getLowerBound}$ function.}
    \label{algo:lb_dffs_par}
\end{algorithm}%
\begin{algorithm}[!t]
    \footnotesize
    \SetAlgoVlined
    \DontPrintSemicolon
    \LinesNumbered
    \KwInput{$f$, $\underline{\lambda}$, $\overline{\lambda}$, $c$, $W_R = [w'_1, \dots, w'_r], k$}
    \KwInOut{$lb$}
    \If{$lb \leq k$}
    {
        $L_f \gets 0$ \tcp*{Only one thread}
        $\mathit{threadsBarrier}()$\;
        $\mathit{tIdx} \gets \mathit{getThreadIndex}()$\;
        $\lambda \gets \underline{\lambda} + \mathit{tIdx}$\;
        \If{$\lambda \leq \overline{\lambda}$}
        {
            $sum \gets 0$ \tcp*{Calculate $L_1$-like lower bound}
            \For{$w \in W_R$}
            {
                $sum \gets sum  + f(w, \lambda)$\;
            }
            $lb' \gets \left\lceil \frac{sum}{f(c, \lambda)} \right\rceil$\;
            $L_f \gets max(L_f, lb')$ \tcp*{Atomic operation}
        }
        $\mathit{threadsBarrier}()$\;
        $lb \gets max(lb, L_f)$ \tcp*{Only one thread, atomic operation}
    }
    \caption{Pseudocode of the \emph{calcDffLowerBound} kernel.}
    \label{algo:lb_dffs_kernel}
\end{algorithm}
\section{Experiments}
\label{sec:experiments}
This section presents the results of a comparison between propagators that use different lower bounds for the feasibility check. We evaluate our implementation of $L_2$\footnote{Algorithm with linear complexity time.}, \cref{algo:lb_dffs}, \cref{algo:lb_dffs_par}, and the implementation from \cite{Cambazard10} which uses the Arc-Flow based lower bound. We refer to them as \texttt{L2}, \texttt{DFFs-CPU}, \texttt{DFFs-GPU}, \texttt{Arc-Flow}.

We select two classic BPP benchmarks from the literature \cite{Falkenauer96,Scholl97}, and generate new instances similar to the ones proposed in \cite{Castineiras12}. This results in a total of $1,922$ instances organized as follows: 
\begin{description}
    \item[Falkenauer] These instances are divided into two classes, each consisting of 80 instances. The `U' instances contain items with weights uniformly distributed in the range $[20, 100]$, $n \in \set{120, 250, 500, 1000}$ and $c = 150$. The `T' instances are more difficult, characterized by triplets of items that must be packed in the same bin in any optimal solution. For this class, $n \in \{60$, $120$, $249$, $501\}$ and $c = 1000$.
    \item[Scholl] These instances are divided into three sets of 720, 480, and 10 instances. Set 1 contains instances where the item weights are uniformly distributed to expect a number of items per bin not larger than three, $n \in \{50$, $100$, $200$, $500\}$, $c \in \set{100, 120, 150}$. Set 2 contains more difficult instances where the item weights are uniformly distributed to expect between three and nine items per bin, $n \in \set{50, 100, 200, 500}$, $c = 1000$. Set 3 contains hard instances with weights uniformly distributed in the range $[20000, 35000]$, $n = 200$ and $c = 100000$.
    \item[Weibull] These instances are based on the Weibull probability distribution. It can model various distributions found in different problem domains by adjusting the shape parameter $k > 0$ and the scale parameter $\lambda > 0$. We generated 92 sets of weights $W$ with the parameters $n \in \{100, 200\}$, $k \in \{0.5, 0.6, \ldots, 5.0\}$, and $\lambda = 1000$. For each set $W$, we generate 6 instances $(c, W)$ with $c = \sigma \cdot \max(W)$ for $\sigma \in \{1.0, 1.2, \ldots, 2.0\}$. The total number of instances is 552, with capacity ranging between 1300 and 92500.

\end{description}

The resolution procedure that we use is the same as in \cite{Shaw04,Cambazard10}, where a minimum number of bins is established and an attempt to find a solution is made. If such a solution does not exist, the number of bins is increased, and a new attempt is made. All implementations use the \emph{decreasing best fit} search heuristic. In this strategy, the items are considered in descending order of weight and assigned to the first bin within their domain that has the smallest residual capacity sufficient to accommodate the item. 
Additionally, two symmetry-breaking rules are applied on backtracking: first, the bin is removed from the domain of all items of the same size, and second, all the bins with the same load are removed from the domains of these items. Finally, a dominance rule is applied before a choice point: if an item completely fills the remaining capacity of a bin, it is assigned to that bin. 

The implementations \texttt{L2}, \texttt{DFFs-CPU}, and \texttt{DFFs-GPU} include additional techniques. First, another dominance rule is applied before a choice point: if at most one item fits in the residual capacity of a bin, the heaviest among such items is assigned to the bin \cite{Schaus09}. Second, the feasibility check uses all three reductions described in \cite{Dupuis10}. Finally, the symmetry breaking described in \cite{Salem20} is enforced with an additional constraint. We discuss the impact of these techniques later in this section.

The experiments are performed with a time limit of 10 minutes to ensure a reasonable benchmark time. The system used for the tests is equipped with an Intel Core i7-10700K processor, 32 GB of RAM, and an NVIDIA GeForce RTX 3080. It runs Ubuntu 22.04, CUDA 11.8, Open JDK 11.0 and CPLEX 22.1.

\subsection{Results and Analysis}
The analysis focuses on instances solved within the time limit. Table \ref{tab:results} reports, for each approach and benchmark, the number of solved instances, the average time per solved instances, the total time to solve them, and the total number of visited nodes. 

\begin{table}[!t]
\setlength{\aboverulesep}{0pt}
\setlength{\belowrulesep}{0pt}
\setlength{\tabcolsep}{6pt} 
\definecolor{tableYellow}{HTML}{FFDF42}
\definecolor{tableGreen}{HTML}{7ED321}
\centering
\begin{NiceTabular}{@{}ccrrrr@{}}
\CodeBefore
\cellcolor{tableYellow}{2-2,2-3,2-4}
\cellcolor{tableYellow}{5-2,5-3,5-4}
\cellcolor{tableYellow}{9-2,9-3,9-4}
\cellcolor{tableYellow}{13-2,13-3,13-4}
\cellcolor{tableYellow}{16-2,16-3,16-4}
\cellcolor{tableYellow}{20-2,20-3,20-4}
\cellcolor{tableYellow}{24-2,24-3,24-4}
\rectanglecolor{white}{2-1}{25-1}
\Body
Benchmark & Lower Bound & Solved & Avg Time [s] & Tot Time [s] & Nodes   \\ 
\midrule
\Block{4-1}{Falkenauer T}       &   \texttt{L2}          & 38              & \textbf{11}    & 408              & 1244813 \\
                                &   \texttt{DFFs-CPU}    & 58              & 64             & 3677             & 891909  \\
                                &   \texttt{DFFs-GPU}    & 58              & 21             & 1196             & 891909  \\
                                &   \texttt{\textbf{Arc-Flow}}    & \textbf{68}     & 20             & 1327             & 5235 \\
\midrule                                                                                                      
\Block{4-1}{Falkenauer U}       &   \texttt{L2}          & 31              & 20             & 628              & 8091399 \\
                                &   \texttt{DFFs-CPU}    & 59              & 60             & 3568             & 584459  \\
                                &   \texttt{DFFs-GPU}    & 60              & 59             & 3559             & 698451  \\
                                &   \texttt{Arc-Flow}    & \textbf{79}     & \textbf{9}     & 690              & 15990 \\
\midrule                                                                                                      
\Block{4-1}{Scholl 1}           &   \texttt{L2}          & 640             & 7              & 4457            & 34421873 \\
                                &   \texttt{DFFs-CPU}    & 700             & 6              & 4374            & 3397547  \\
                                &   \texttt{DFFs-GPU}    & 703             & 6              & 3985            & 7901777  \\
                                &   \texttt{Arc-Flow}    & \textbf{717}    & \textbf{3}     & 2492            & 115520 \\
\midrule                                                                                                      
\Block{4-1}{Scholl 2}           &   \texttt{L2}          & 336             &  5             & 1620            & 34401378 \\
                                &   \texttt{DFFs-CPU}    & 437             &  9             & 3950            & 596417  \\
                                &   \texttt{DFFs-GPU}    & \textbf{440}    &  \textbf{4}    & 1951            & 2400806  \\
                                &   \texttt{Arc-Flow}    & 436             & 61             & 26800           & 278695 \\
\midrule                                                                                                      
\Block{4-1}{Scholl 3}           &   \texttt{L2}          & 0               & \textendash    & \textendash     & \textendash \\
                                &   \texttt{DFFs-CPU}    & 3               &  316           & 947             & 8221  \\
                                &   \texttt{DFFs-GPU}    & \textbf{3}      & \textbf{1}     & 3               & 8221  \\
                                &   \texttt{Arc-Flow}    & 0               & \textendash    & \textendash     & \textendash \\
\midrule                                                                                                     
\Block{4-1}{Weibull}           &   \texttt{L2}           & 375             &  70            & 26301           & 55122362 \\
                                &   \texttt{DFFs-CPU}    & 402             &  9             & 3432            & 722730  \\
                                &   \texttt{DFFs-GPU}    & \textbf{418}    &  \textbf{9}    & 3652            & 35003432  \\
                                &   \texttt{Arc-Flow}    & 292             &  95            & 28022           & 11437 \\
\end{NiceTabular}
    \caption{Statistics for the solved instances of different lower bound methods.}
    \label{tab:results}
\end{table}
\begin{figure}[!t]
\centering
\includegraphics[width=0.49\linewidth]{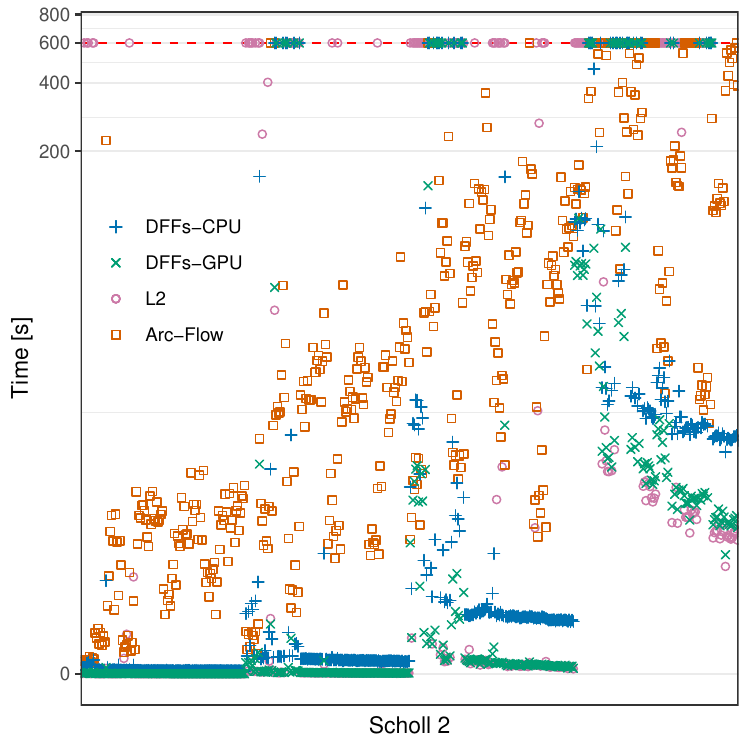}
\includegraphics[width=0.49\linewidth]{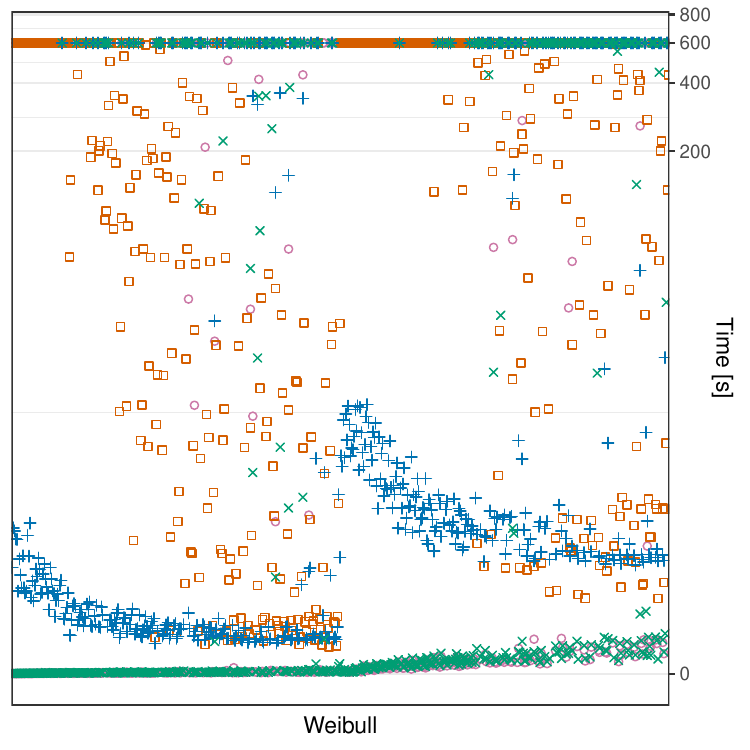}
\caption{Solve time for the Scholl 2 and Weibull instances. Note the logarithmic scale in the time axe. The \texttt{DFFs-GPU} points are colored in green.}
\label{fig:solve_time}
\end{figure}

The Falkenauer T instances highlight the contrast between fast but weak and slow but strong pruning. $\texttt{L2}$ quickly solved almost half of the instances, while \texttt{Arc-Flow} solved 85\% of them, taking on average twice the time per instance. The DFFs-based approaches fall in the middle, with the GPU-accelerated implementation being three times faster than the sequential version.

In the Falkenauer U and Scholl 1 instances, \texttt{Arc-Flow} demonstrated a good balance between speed and strength, solving almost all instances in a small amount of time. The DFFs-based approaches follow by number of solved instances, and $L_2$ comes last. Notice how the gap between \texttt{Arc-Flow} and the DFFs-based approaches shrinks from small (Falkenauer U) to larger (Scholl 1) bin capacities. The additional instances solved by \texttt{DFFs-GPU} compared to \texttt{DFFs-CPU} accounted for 15\% (Falkenauer U) and 29\% (Scholl 1) of the total solving time.

The results for the Scholl 2 instances (see \cref{fig:solve_time}) show that \texttt{DFFs-GPU} is a competitive approach, solving a larger number of instances with the minimum average time. Closely following is \texttt{DFFs-CPU}, which is twice slower compared to the GPU version, and \texttt{Arc-Flow} that requires significantly more time. Finally, $L_2$ comes last, with a good average resolution time but fewer solved instances.

The Scholl 3 instances, characterized by their huge capacities, highlight the convenient trade-off between tight bounds and computational speed offered by DFFs. These approaches are the only ones able to solve some instances, with the GPU-accelerated implementation showing remarkable speedups. 

The results obtained from the Weibull instances (see \cref{fig:solve_time}) confirm the effectiveness of the DFFs-based approaches, especially for large bin capacity. \texttt{DFFs-GPU} solved more instances than it sequential counterpart, which accounts for 96\% of the solving time. In third place is $L_2$ with a significantly larger average solving time, and last comes \texttt{Arc-Flow} with significantly fewer solved instances.

In conclusion, the DFFs-based approaches offer an interesting tradeoff between pruning strength and computational speed that becomes more valuable as the bin capacity increases. The speedups provided by the GPU depend on both the capacity of the bins and the characteristics of the instance. As the calculation of the lower bound is the only GPU-accelerated operation, the benefits are proportional to the number of times the feasibility check is performed. This count can be significantly smaller than the number of propagator calls, since failures can occur earlier in the knapsack reasoning.

To assess the impact of the different optimizations we employed, an ablation study has been conducted on the instances solved by \texttt{DFFs-GPU} in less than 60 seconds. The results are presented in \cref{tab:opt_diff}, where each entry represents a version of \texttt{DFFs-GPU} with a disabled optimization. The most effective techniques are the dominance rule \cite{Shaw04} and symmetry breaking \cite{Salem20}. In general, we strongly encourage the use of the latter since it is implementable as a standalone constraint and applicable to variations of the BPP with precedences or conflicts.

\begin{table}[t]
    \setlength{\tabcolsep}{6pt} 
    \centering
    \begin{tabular}{@{}crrr@{}}
    Version           & Solved & Time [s] & Nodes    
    \\ \midrule
    \texttt{DFFs-GPU}          & 1631   & 1847     & 4653344  \\
    \texttt{DFFs-GPU-NoDom}    & 1598   & 2515     & 16420684 \\
    \texttt{DFFs-GPU-NoAltRed} & 1623   & 2597     & 12986578 \\
    \texttt{DFFs-GPU-NoSymBrk} & 1591   & 5970     & 53424505
    \end{tabular}
        \caption{Statistics for \texttt{DFFs-GPU} without optimizations.}
    \label{tab:opt_diff}
\end{table}





\section{Conclusions and Future works}
\label{sec:conslusions}
This paper discusses the Bin Packing Problem, presenting a feasibility check using different lower bounds derived from Dual Feasible Functions. While these lower bounds may not be the fastest or the tightest, the substantial parallelism offered by modern GPUs changes this position, making the approach effective, particularly for large problem  instances.

This work raises several research questions that could be explored in future studies. From an analytical standpoint, it would be interesting to identify functions that lead to tight bounds in cases where the current ones fall short. On a practical note, a valuable extension would be to explore the effectiveness of multidimensional Dual Feasible Functions \cite{Alves14} on the Multidimensional/Vector Bin Packing Problem.

\section*{Acknowledgements}
We would like to thank François Clautiaux and Maxence Delorme for their valuable feedbacks, Hadrien Cambazard for generously providing the original Arc-Flow implementation, and Jürgen Rietz for sharing insightful perspectives on Dual Feasible Functions.

\bibliographystyle{splncs04}
\bibliography{main}

\end{document}